\documentclass[twocolumn,fleqn,usenatbib]{openjournal}
\usepackage{hyperref}
\usepackage{bm}
\usepackage{multirow}
\usepackage{gensymb}
\usepackage{enumitem}
\usepackage{tabularx}
\usepackage[flushleft]{threeparttable}
\usepackage[]{color}
\usepackage{calc}

\usepackage{float}
\usepackage{breqn}
\usepackage{amsmath}

\usepackage{amsfonts,amsmath,amssymb,esint}
\usepackage{graphicx}
\usepackage{color}
\usepackage{xcolor}
\usepackage{url}

\begin{document}

\title{A Machine Learning Approach to Coupling Photoionization \& Multi-Frequency Radiation Hydrodynamics}

\author{
Sergei Dyda$^{1,\ast}$ \&
Randall C. Dannen$^{2,3}$
}

\thanks{$^\ast$Corresponding author: sd449@cornell.edu}

\affiliation{
$^1$Department of Physics \& Astronomy, University of Alabama,
Gallalee Hall, 514 University Blvd, Tuscaloosa, AL 35401, USA
}

\affiliation{
$^2$Department of Physics \& Astronomy, University of Nevada, Las Vegas,
4505 S. Maryland Pkwy, Las Vegas, NV 89154-4002, USA
}

\affiliation{
$^3$Nevada Center for Astrophysics, University of Nevada, Las Vegas,
4505 S. Maryland Pkwy, Las Vegas, NV 89154, USA
}

%%%%%%%%%%%%%%%%%%%%%%%%%%%%%%%%%%%%%%%%%%
\begin{abstract}
We use machine learning (ML) to couple photoionization microphysics to radiation hydrodynamic simulations of line-driven winds. ML models are trained on outputs from the photoionization code \textsc{XSTAR} to predict the line-driving force multiplier as a function of gas ionization, optical depth, and the blackbody temperature of the irradiating spectral energy distribution (SED). We implement the trained models within \textsc{Athena++}, enabling rapid runtime evaluation of the radiative acceleration without directly solving the photoionization problem during the hydrodynamic evolution. We find that random forest models provide the highest accuracy and reproduce the force multipliers over a three-dimensional parameter space more accurately, and with less training data, than traditional precomputed tables using trilinear interpolation. Because the framework treats the force multiplier as a function of the evolving local SED, it can be generalized to more complex radiation fields and arbitrary spectral shapes. More broadly, this work demonstrates how ML can serve as a computational framework for embedding expensive microphysical calculations within large-scale dynamical simulations.

\end{abstract}

\maketitle

%%%%%%%%%%%%%%%%%%%%%%%%%%%%%%%%%%%%%%%%%%%%%%%%%%%%%%%%%%%%%%%%%%%%%
\section{Introduction}
\label{sec:introduction}

Large-scale gas dynamics and small-scale microphysics are often modeled using fundamentally different numerical approaches. Hydrodynamics and magnetohydrodynamics codes evolve the bulk motion of gas by solving the conservation equations of fluid dynamics, while specialized microphysics solvers compute local quantities such as radiative opacities, heating and cooling rates, chemical abundances and nuclear reaction networks. These small-scale processes strongly influence the dynamics by setting the local forces and thermodynamic state of the gas, while the evolving density, temperature, velocity gradients, and radiation field simultaneously affect the microphysical state. A fully self-consistent treatment therefore requires both calculations to be coupled throughout the simulation. In practice, however, directly evaluating expensive microphysics calculations at every cell and timestep is often computationally prohibitive, particularly in multidimensional simulations. Alternatively, storing precomputed microphysics tables rapidly becomes expensive to construct, store, and interpolate as the number of parameters grows. 
Machine learning (ML) provides a promising framework for coupling expensive microphysical calculations to large-scale dynamical simulations by training surrogate models on precomputed microphysics data and evaluating them during the hydrodynamic evolution.

Radiation-driven winds are a ubiquitous mechanism for mass loss across astrophysics, appearing in systems ranging from massive stars, cataclysmic variables (CVs) and active galactic nuclei (AGN). Such winds provide a particularly important example of this coupling problem, since the radiative force depends sensitively on the local ionization state and line opacities of the gas. Accurately modeling these outflows therefore requires connecting detailed photoionization calculations to the large-scale hydrodynamic evolution of the flow.

The acceleration experienced by the gas irradiated by a frequency-integrated radiative flux $\mathbf{F}_{\rm{rad}}$ 
\begin{equation}
    \mathbf{a} = \frac{\mathbf{F}_{\rm{rad}}}{c} \kappa_{\rm{es}} M_{\rm{total}},
\end{equation}
where $c$ is the speed of light, $\kappa_{\rm{es}}$ the opacity due to electron scattering and $M_{\rm{total}}$ is the force multiplier (\cite{CAK1975}, hereafter CAK). The radiative flux sets the available momentum budget (via $\mathbf{F}/c$), while the force multiplier quantifies how efficient this momentum transfer is relative to electron scattering. Correctly simulating line driven outflows thus depends on accurately computing the radiation flux and the force multiplier. Crucially however, both the source of momentum (via $\mathbf{F}/c$) and the strength of the coupling (via $M_{\rm{total}}$) depend on the local irradiating spectral energy distribution (SED), which in many systems is varying with time.  

Bulk gas motions in radiation-driven outflows are typically evolved using hydrodynamics (or radiation-hydrodynamics), which solves the conservation equations for mass, momentum, and energy under the influence of forces such as gravity and radiative acceleration. The radiative force itself depends on the force multiplier, which must be computed from detailed atomic physics—usually via a photoionization code that determines the ionization balance and line opacities for the local radiation field. Because the force multiplier varies with quantities like density, temperature, and radiation spectrum, it ideally needs to be updated throughout the flow. Running a full hydrodynamic simulation while simultaneously calling a photoionization solver at each location and timestep is very computationally expensive, making this coupled approach extremely challenging in practice. In this work, we use ML to emulate photoionization force multiplier calculations and couple them to radiation hydrodynamic simulations.

A variety of approaches have been tried to include photoionization physics in radiation driven wind simulations including analytic fits \cite{Stevens1990,PK04,Nomura2016,Nomura2017,Nomura2020} and pre-computed lookup tables \cite{Dyda2025b}. These approaches generally assumed a fixed irradiating SED for computing the force multiplier, an approximation that does not hold in these highly time-dependent systems. A more recent approach involved parametrizing the SED as a blackbody, and interpolating across different blackbody temperatures using fitting functions \cite{Dyda2025a}. This allowed for on the fly calculations of the force multiplier during hydrodynamical simulations though the model accuracy struggled in certain parts of the parameter space and this method could not generalize to more general SEDs. 

Simulations of radiation-driven winds in CVs \citep{Pereyra2003, Mosallanezhad2025} and AGN \citep{Matthews2025, Scepi2026} have attempted to evolve the radiation field and hydrodynamics simultaneously. These models evolve the gas dynamics using \textsc{PLUTO}, while the radiation field and force multipliers are computed using Monte-Carlo radiative transfer with \textsc{SIROCCO} and an extensive line list of $\approx 450,000$ transitions \citep{Parkin2013}. Because these calculations are very computationally expensive, the radiation field and force multipliers are updated intermittently rather than at every hydrodynamic timestep. Alternatively, Monte-Carlo radiation transfer can compute the radiation force on the gas directly, without resorting to a force multiplier prescription \cite{2015MNRAS.453.3120N}.

ML surrogate models have recently been applied to accelerate expensive microphysical calculations in astrophysics, including heating and cooling rates in photoionized gas \citep{2024MNRAS.528..255R,2025OJAp....8E..76R}, the evolution of chemical species in the ISM \citep{2023MNRAS.518.5718B,2024A&A...684A.203B}, nuclear reaction networks \citep{2025ApJ...990..105Z}, non-LTE population levels in stellar atmospheres \citep{2022A&A...658A.182C} and non-local thermodynamic equilibrium opacity calculations in radiation hydrodynamics \citep{2020PhPl...27e2707K}. These approaches replace computationally expensive microphysics solvers with ML models trained on precomputed datasets, allowing the microphysical quantities to be rapidly evaluated during the dynamical evolution. Such studies demonstrate that ML can efficiently reproduce complex microphysical calculations while remaining computationally compatible with multidimensional hydrodynamic and radiation-hydrodynamic simulations.  

In this work, we address the challenge of dynamically coupling evolving SED-dependent microphysics to hydro simulations using ML. We demonstrate this by training ML models on precomputed photoionization datasets to predict the force multiplier directly from the radiation and hydrodynamic variables, including a parameterization of the SED. These trained models are queried within Athena++, enabling rapid, on-the-fly evaluation of the force multiplier during the hydrodynamic evolution. This approach allows the radiation field and gas dynamics to be dynamically coupled while making the problem computationally tractable.

The paper is organized as follows. In Section \ref{sec:photoionization}, we describe the photoionization calculations used to generate the training data. Section \ref{sec:machinelearning} presents the ML framework to train and benchmark the models. In Section \ref{sec:hydro}, we incorporate the trained model into hydrodynamic simulations, demonstrating how ML can be used to evaluate force multipliers using the self-consistent SED produced within \textsc{Athena++}. Finally, Section \ref{sec:discussion} discusses the implications and limitations of the method, along with future applications to more realistic line-driven wind calculations and to other physical systems where expensive microphysical calculations must be coupled to large-scale dynamical simulations.

\section{Photoionization Calculations and ML Training Data}
\label{sec:photoionization}

In line-driven winds, momentum is transferred from the radiation field to the gas via bound-bound atomic transitions. Following the Sobolev approximation, the contribution from an individual spectral line depends on both the local radiation field and the optical depth of the transition.

We employ the photoionization code \textsc{XSTAR} to compute a grid of photoionization models used to determine the radiative force multipliers incorporated into our hydrodynamic simulations. We quantify the state of the gas via the ionization parameter
\begin{equation}
    \xi = \frac{4 \pi F}{n_H},
\end{equation}
the ratio of ionizing flux $F$ and the gas number density $n_H$ and the optical depth parameter 
\begin{equation}
    t = \sigma_e \rho l_{\rm{Sob}}, 
\end{equation}
where $\sigma_e$ is the electron scattering opacity, $\rho$ the gas density and 
\begin{equation}
    l_{\rm{Sob}} = v_{\rm{th}} / |dv/dl|
\end{equation}
is the Sobolev length with $dv/dl$ the velocity gradient of the flow along the path of the radiation. The irradiating SED is assumed to be a blackbody with temperature $T_{bb}$. 

The version of \textsc{XSTAR} used in this work includes a line list containing more than $2 \times 10^6$ lines (\cite{Dannen19}, hereafter D19. See \cite{Abbott1982} for the earliest studies using line lists). We assume same atomic abundances as D19. 

The force per unit mass due to an individual line at frequency $\nu_0$ can be computed as 
\begin{equation}
    f_L = \frac{F_{\nu} \Delta \nu_D}{c} \frac{\kappa_L}{\tau_L} \left( 1 - e^{-\tau_L}\right),
\label{eq:aline}    
\end{equation}
where $\kappa_L$ is the line's opacity, $\tau_L$ the optical depth, $\Delta \nu_D = \nu_0 v_{\rm{th}}/c$ the thermal Doppler width, with $v_{\rm{th}}$ the thermal speed of the ion corresponding the the atomic line in question and $F_{\nu}$ is the specific flux. 

For a single line, the opacity due to stimulated emission, in units of $\rm{cm^2 g^{-1}}$ is
\begin{equation}
    \kappa_L = \frac{\pi e^2}{m_e c} gf \frac{N_L/g_L - N_U/g_U}{\rho \Delta \nu_D},
\end{equation}
where $gf$ is the oscillator strength and $N_L$ ($N_U$) and $g_L$ ($g_U$) are the occupation number densities and weights of the lower (upper) levels respectively. Rewriting the optical depth as
\begin{equation}
    \tau_L = \eta t,
\end{equation}
where $\eta = \kappa_L / \sigma_e$, we can express the total acceleration as a sum over the individual line contributions (\ref{eq:aline}), 
\begin{equation}
    \mathbf{a}_{\rm{tot}} = \frac{\sigma_e}{c} \mathbf{F} M_{\rm{tot}}(t,\xi),
\end{equation}
where the force multiplier is 
\begin{equation}
    M_{\rm{tot}}(t,\xi) = \sum_{\rm{lines}} \frac{F_{\nu} \Delta \nu_D}{\sigma_e F} \frac{\kappa_L}{\tau_L} \left( 1 - e^{-\tau_L}\right).
    \label{eq:Mtotal}
\end{equation}
Different decompositions of the force multiplier are useful for different applications, including reduced-order ML models, spectral diagnostics, and coupling to multifrequency radiation hydrodynamics. For example, we can define the force multiplier for a frequency band $\nu$ 
\begin{equation}
M_{\nu}(t,\xi) = \sum_{\substack{\rm lines, \\ \nu}} 
\frac{F_{\nu} \Delta \nu_D}{\sigma_e F} 
\frac{\kappa_L}{\tau_L} \left( 1 - e^{-\tau_L}\right),
\label{eq:Mband}
\end{equation}
where the sum is carried out over all lines in the relevant frequency range. Alternatively, we may group the contributions over all line transitions available to a certain ion
\begin{equation}
    M_{\rm{ion}}(t,\xi) = \sum_{\substack{\rm{lines,} \\ \rm{ion}}} \frac{F_{\nu} \Delta \nu_D}{\sigma_e F} \frac{\kappa_L}{\tau_L} \left( 1 - e^{-\tau_L}\right).
        \label{eq:Mion}
\end{equation}

For a wide range of optical depth parameter $-8 \leq \log t \leq 1$, photoionization parameters $-2 \leq \log \xi \leq 5$ and blackbody temperatures $10^4 \ \rm{K} \leq T_{bb} \leq 10^{6} \ \rm{K}$ we compute the force multiplier (see D19, for a full description of our methods) in different formats: the total (\ref{eq:Mtotal}), by band (\ref{eq:Mband}) (for IR, O, UV and X bands) and by ion (\ref{eq:Mion}) for a list of 496 ions, corresponding to all possible ionization states for elements $H$ through $Ga$. We show a selection of this data in Appendix \ref{sec:appendix}. These multidimensional tables form the training and testing datasets for the ML models described in the following section, where the goal is to efficiently reproduce the force multipliers across the full $(t, \xi, T_{\rm{bb}})$ parameter space.

\section{Machine Learning}
\label{sec:machinelearning}

The dataset was constructed from force multiplier tables computed for 11 blackbody spectra spanning temperatures $10^4 \ \rm{K} \leq T_{bb} \leq 10^6 \ \rm{K}$, ionization parameter $-2 \  \leq \log \xi \leq 5$ and optical depth parameter $-8 \ \leq \log t \leq 1$. This produced a three-dimensional parameter space containing $N = 177100$ samples. The training data therefore sample the $(t,\xi,T_{\rm{bb}})$ parameter space approximately uniformly in logarithmic coordinates.  We randomly assigned 80\% of the full data set to training and the remaining 20\% to testing model accuracy. From this training data, we would typically use $N = 80k$ data points which we found yielded reasonable model accuracy (see below). 

Our fiducial model was a random forest regressor using the \texttt{sklearn} module part of the scikit-learn\footnote{\url{https://scikit-learn.org}} machine learning library available for the \texttt{Python} programming language \citep{scikit-learn}. The model was implemented using a standard ensemble of decision trees trained via bootstrap aggregation. The model consisted of 200 trees, each grown to full depth (i.e., no explicit maximum depth constraint) with a minimum of one sample per leaf, allowing the trees to capture fine structure in the data. A fixed random seed was used to ensure reproducibility. The model natively supports multi-output regression, enabling simultaneous prediction of the target quantities without additional wrappers. We explored variations in the number of trees ($\sim$100–1000), minimum samples per leaf (1–20), and maximum tree depth, and found that the chosen configuration provided a robust balance between accuracy and computational efficiency.

The training output was a model predicting $M_{IR}, M_O, M_{UV}$ and $M_{X}$ - the force multiplier in 4 frequency bands corresponding to infrared, optical, ultraviolet and X-rays respectively.

To score model performance, we compared the model prediction $M_{ML}$ to the \textsc{XSTAR} value $M_{XSTAR}$ for all points $N = 35420$ in the testing set via   
\begin{equation}
   \Delta_{\rm{dex}} = |\log M_{ML} - \log M_{\rm{XSTAR}}|
   \label{eq:error}
\end{equation} 

With this metric, we consider points with $\Delta_{\rm{dex}} \leq 0.01$ to be sufficiently accurate, corresponding to $\leq 2.3\%$ relative error. By contrast, we consider an error of $0.01 \leq \Delta_{\rm{dex}} \leq 0.05$ as acceptable - this is a relative error of 12\% on the upper end. Finally, we consider $\Delta_{dex} \geq 0.05$ as inaccurate.

Further, we only include points where the force multiplier is greater than a threshold $M_{\nu} > 10^{-2}$, since below this threshold acceleration due to lines is negligible compared to electron scattering.  Our strategy is to find models which maximize the ratio of accurate points while simultaneously minimizing the number of inaccurate points to zero. 

We examined the distribution in phase space of the inaccurate $\Delta_{dex}> 0.05$ points and found them to be mostly uniformly distributed across all the values of the input parameters, though we note slight preference near the edges of the parameter space. This is somewhat expected, as this ML model is better suited for interpolation, rather than extrapolation. We conclude that any use of the model should be restricted to the region of parameter space over which it was trained.

\paragraph*{Training Set Size}

 \begin{figure}
    \centering
        \includegraphics[width=\linewidth]{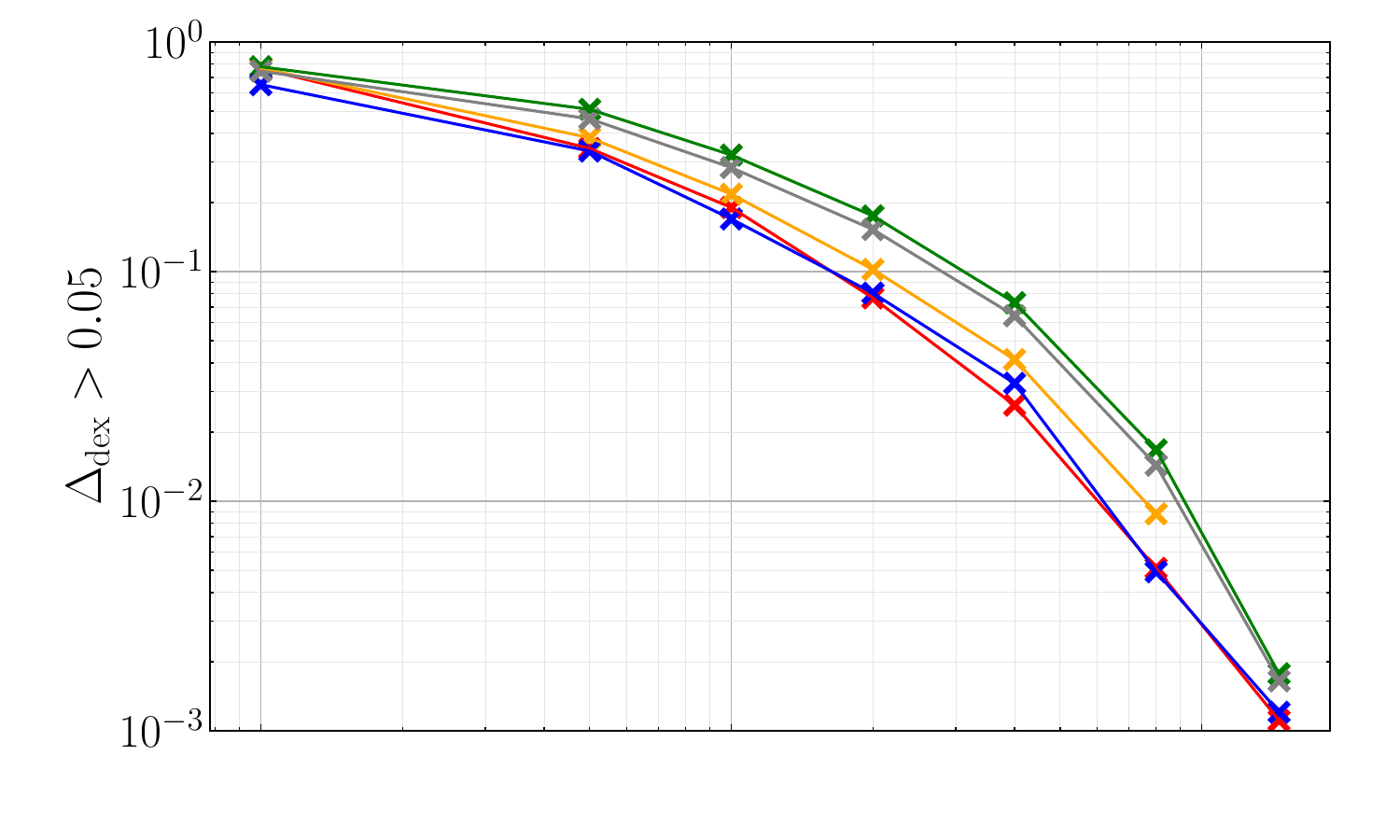}\\[-4mm]
        \includegraphics[width=\linewidth]{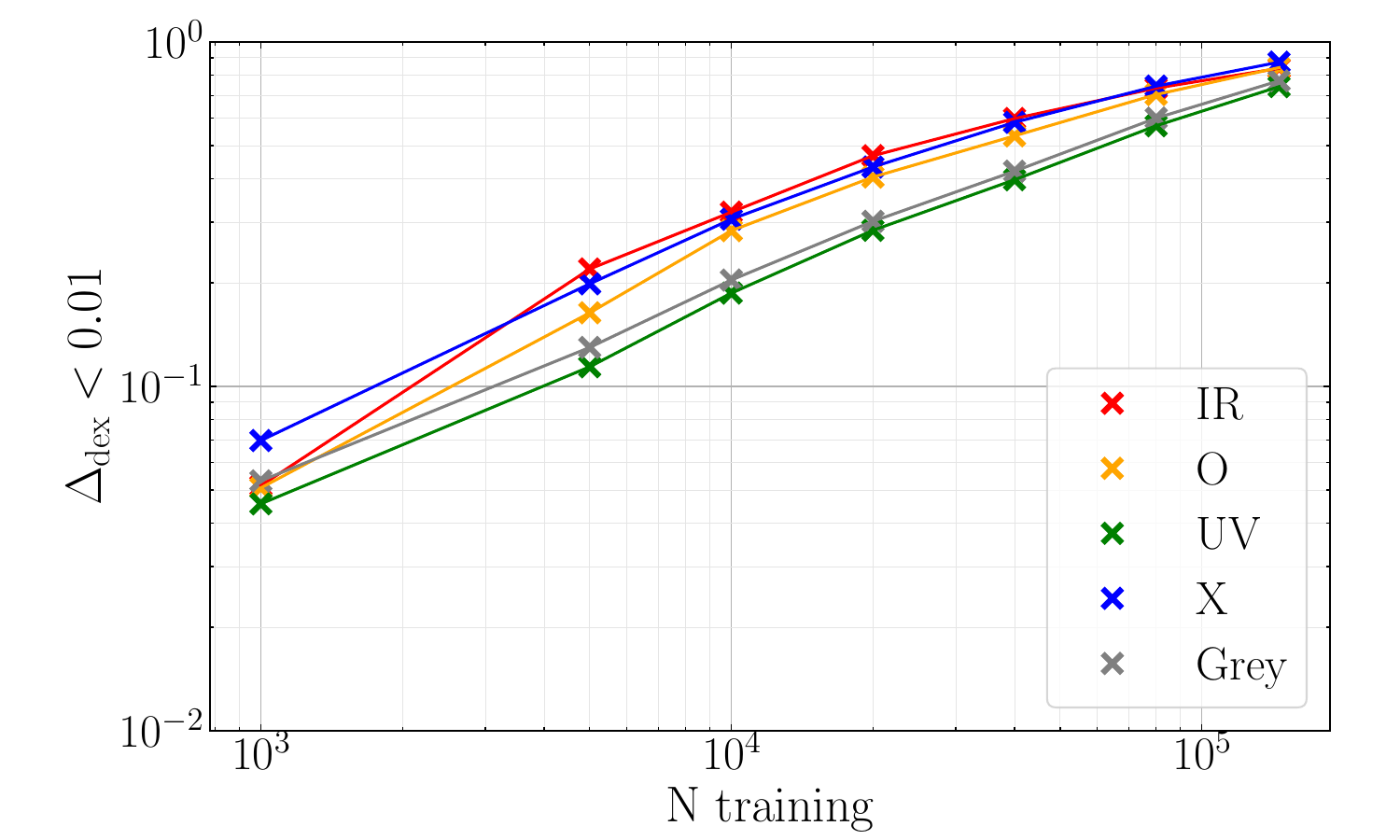}
    \caption{\textit{Top -} Fraction of force multiplier points below the $\Delta_{\rm{dex}} \geq 0.05 $ accuracy threshold as a function of points in the training set for different frequency bands and their total, denoted by ``grey''. \textit{Bottom -} Fraction of force multiplier points exceeding $\Delta_{\rm{dex}} \leq 0.01 $ accuracy threshold as a function of points in the training set for different frequency bands.}
    \label{fig:MLConverge}
\end{figure}

We experimented with convergence of model performance as a function of training set size. In the upper panel of Fig \ref{fig:MLConverge} we show how the number of ``inaccurate'' predictions with a error $\Delta_{\rm{dex}} > 0.05$ decreases as a function of the size of the training set N for each band, as well as the total force multiplier denoted by ``grey''. With $N = 80k$, approximately 1\% of data points have an error with $\Delta_{\rm{dex}} > 0.05$. Using the full training set, $N \approx 142k$ points this number reduces to approximately $0.1 \%$. The lower panel of Fig \ref{fig:MLConverge} shows the fraction of ``accurate'' model points, with $\Delta_{\rm{dex}} < 0.01$. With $N=80k$, roughly 60\% of data points are within this accuracy, though using the full training set can achieve nearly 70\% of points reaching this benchmark. We find $N = 80k$ training points provides an acceptable level of accuracy to the ML model. We therefore use it as the baseline size of our training sets for other model studies. 

\paragraph*{Model Inputs}
Using a baseline training set with $N=80k$ points, we used different inputs to train the model. We tried parametrizing the SED by the blackbody temperature, $T_{\rm{bb}}$, its logarithm $\log T_{\rm{bb}}$ and by the fraction of UV photons $f_{UV}$. We found no difference between using the linear or logarithmic value of the blackbody temperature. We note that the training set covers the relevant temperature range approximately logarithmically, and this is probably sufficient to capture changes in behaviour. Interestingly, using the UV fraction does not perform as well - approximately 7 \% of points have $\Delta_{\rm{dex}} > 0.1$. Because the UV fraction varies rapidly with temperature near the extrema of our temperature range but is relatively insensitive to temperature at intermediate values, it performs poorly as an input parameter. This behavior demonstrates that physically motivated input parameters are not necessarily optimal for ML regression if they compress or distort important regions of the parameter space. We therefore retain $T_{\rm bb}$ as the fiducial SED input parameter for the remainder of this work. 

\paragraph*{Model Outputs}

 \begin{figure}
    \centering
        \includegraphics[width=\linewidth]{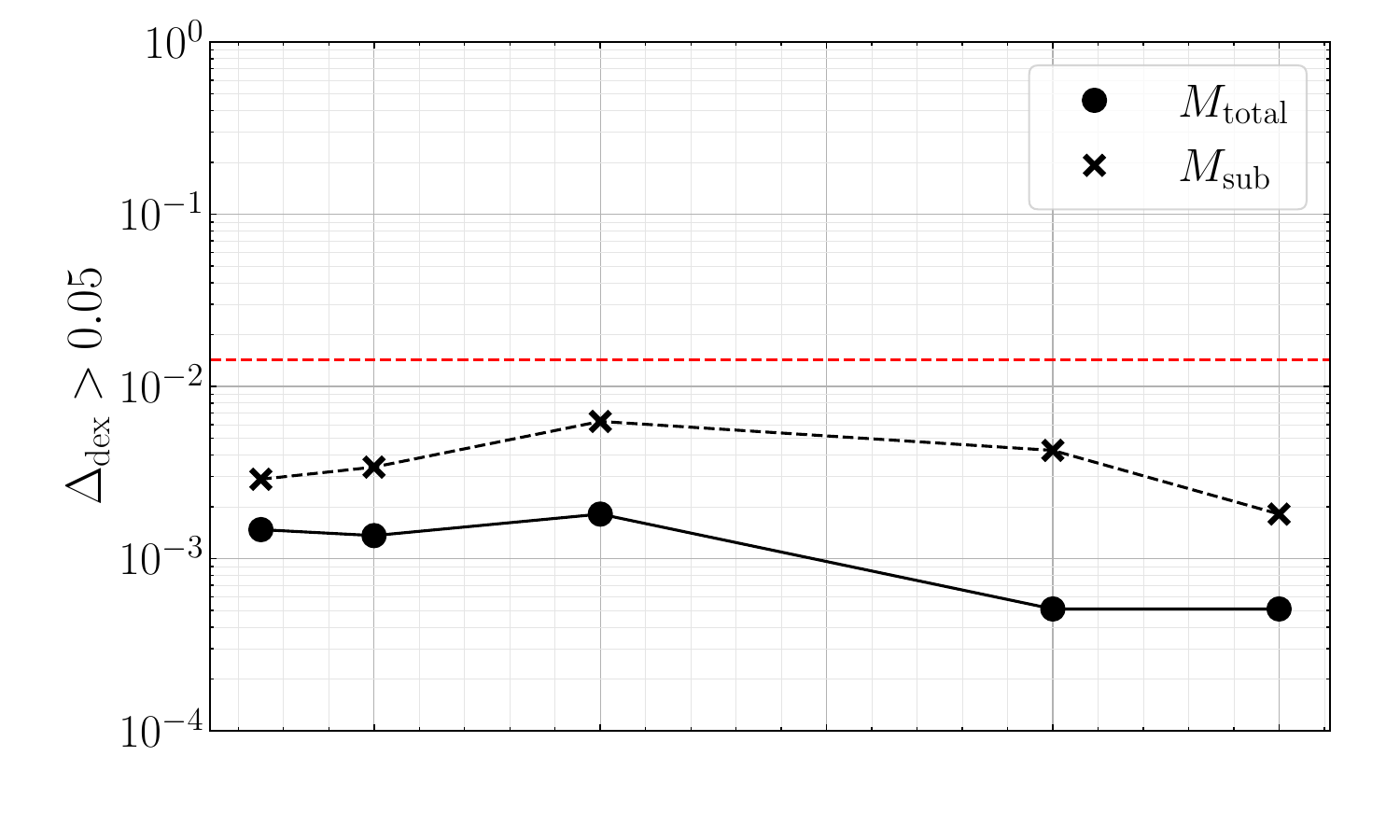}\\[-4mm]
        \includegraphics[width=\linewidth]{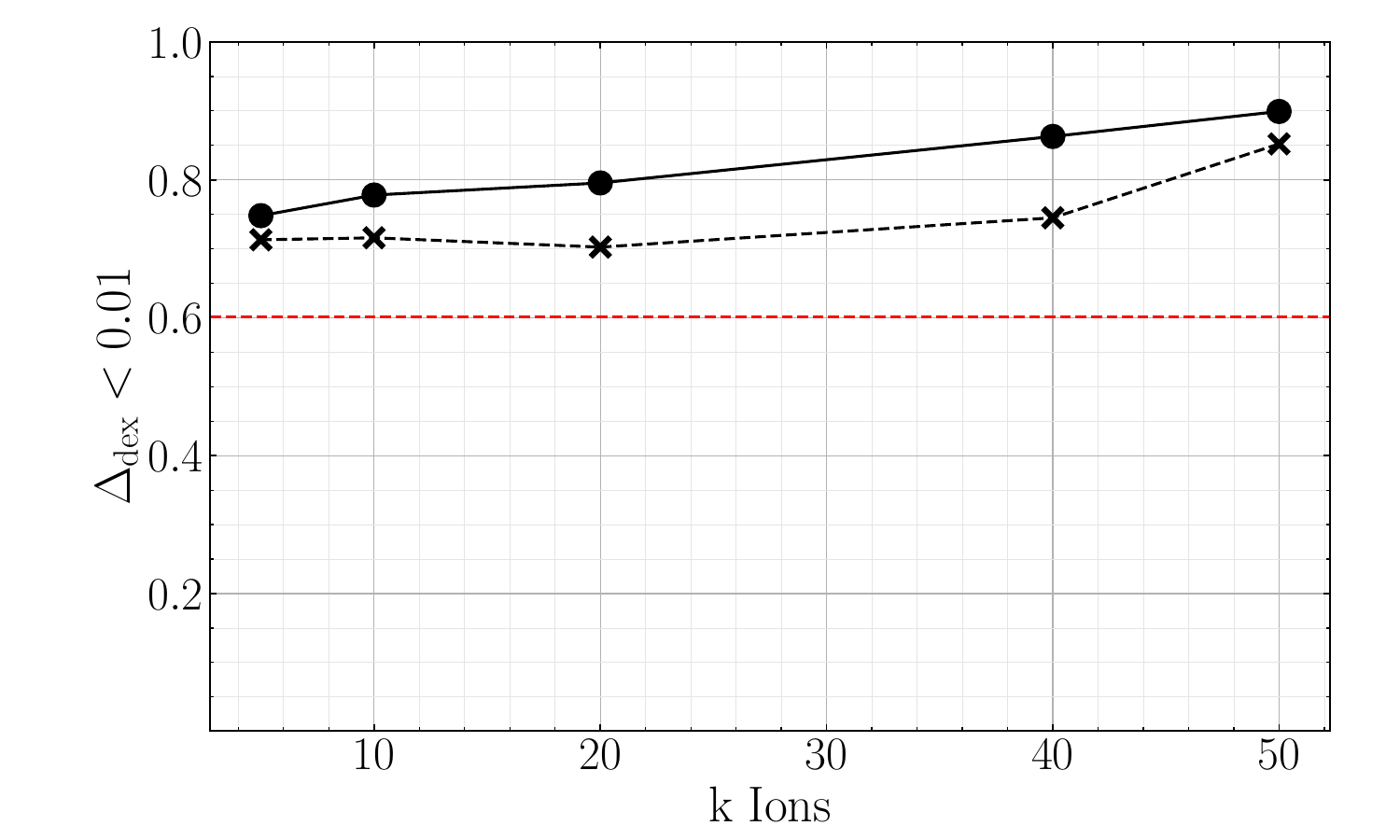}
    \caption{\textit{Top --} Fraction of force multiplier predictions with $\Delta_{\rm dex} \geq 0.05$ as a function of the number of individually modeled ions, $k$. \textit{Bottom --} Fraction of force multiplier predictions with $\Delta_{\rm dex} \leq 0.01$ as a function of $k$. The fiducial four-band model is shown for comparison.}
    \label{fig:MLIonConverge}
\end{figure}

We also explored training the model on force multiplier contributions from individual ions, $M_{\rm ion}$ (see equation (\ref{eq:Mion}), rather than on the band-limited force multipliers. Using the baseline training set of $N=80k$ points, we selected the $k$ ions with the largest total contribution to the force multiplier across the simulation domain from the 496 available ions. We considered models with $5 \leq k \leq 50$. Contributions from all remaining ions were combined into a single ``other'' category. The model was therefore trained on $k+1$ outputs, which were summed to reconstruct $M_{\rm total}$ for each point in the testing set. The reconstructed values were then compared against the corresponding \textsc{XSTAR} values using the same error metric (\ref{eq:error}) as above.

Figure~\ref{fig:MLIonConverge} shows the accuracy of the ion-based models as a function of $k$. The circles show the accuracy of the total force multiplier and the crosses of the worst performing individual ion.  The upper panel shows the fraction of inaccurate predictions, defined by $\Delta_{\rm dex} \geq 0.05$, while the lower panel shows the fraction of accurate predictions, defined by $\Delta_{\rm dex} \leq 0.01$. For comparison,the dashed red line shows the performance of the fiducial four-band model.

These results show an increase by $\sim 10\%$ in accurate points and a decrease by a factor of $\sim 10$ in the number of inaccurate points. Performance does not depend strongly on the number of ionic contributions, for $k \gtrsim 5$. The ``other'' category is most difficult to model when it contains a mixture of ions that still contribute significantly to the total force multiplier. As $k$ increases, the remaining ions contribute less to $M_{\rm total}$, and the error associated with the ``other'' category decreases. However, modeling a large number of individual ions becomes increasingly noisy because the identities of the most important ions vary across the parameter space. For this reason, we do not extend the ion-based models beyond $k=50$. A more accurate ion-based approach would likely require a local or adaptive selection of the dominant ions rather than a single globally ranked ion list. Further, this method requires calling $k$ different models to evaluate the total force multiplier. For this reason, we continue to use the band based approach as the fiducial method.

\paragraph*{Alternative ML Models}

To assess alternative ML approaches, we implemented both gradient boosting and multilayer perceptron (MLP) models using the same training set, with $N=80k$ points, and the same testing set as the random forest baseline. The gradient boosting model was constructed using a histogram-based implementation with a squared-error loss function, a learning rate of 0.03, and 1000 boosting iterations. The model used trees with a maximum of 63 leaf nodes and a minimum of 10 samples per leaf, with no additional $\ell_2$ regularization. These parameters were chosen to provide sufficient model flexibility while maintaining stable training. However, the gradient boosting model performed substantially worse than the random forest baseline, with approximately 10\% of test points having errors $\Delta_{\rm dex} > 0.1$.

For the MLP model, we used a feedforward neural network implemented as part of a pipeline with input standardization via a \texttt{StandardScaler}. The network consisted of two hidden layers with 128 neurons each, ReLU activation functions, and was trained using the Adam optimizer. The model was trained for up to 1000 iterations with an initial learning rate of $10^{-3}$ and adaptive learning rate scheduling. Early stopping was enabled using a 10\% validation split and a patience of 20 iterations. A small $\ell_2$ regularization term, $\alpha = 10^{-4}$, was included to reduce overfitting. To handle multiple outputs, the network was wrapped in a \texttt{MultiOutputRegressor}. Model performance was evaluated using the same testing set and logarithmic error metric defined in Eq.~(\ref{eq:error}).

The MLP model performed worse than both the random forest and gradient boosting models, with more than 20\% of test points having errors $\Delta_{\rm dex} > 0.1$. Overall, the superior performance of the random forest model, together with its computational efficiency and sufficient accuracy for our purposes, led us to adopt random forests as the fiducial ML method for the remainder of this study.

\paragraph*{Trilinear Interpolation Test}

\begin{figure}
    \centering
        \includegraphics[width=\linewidth]{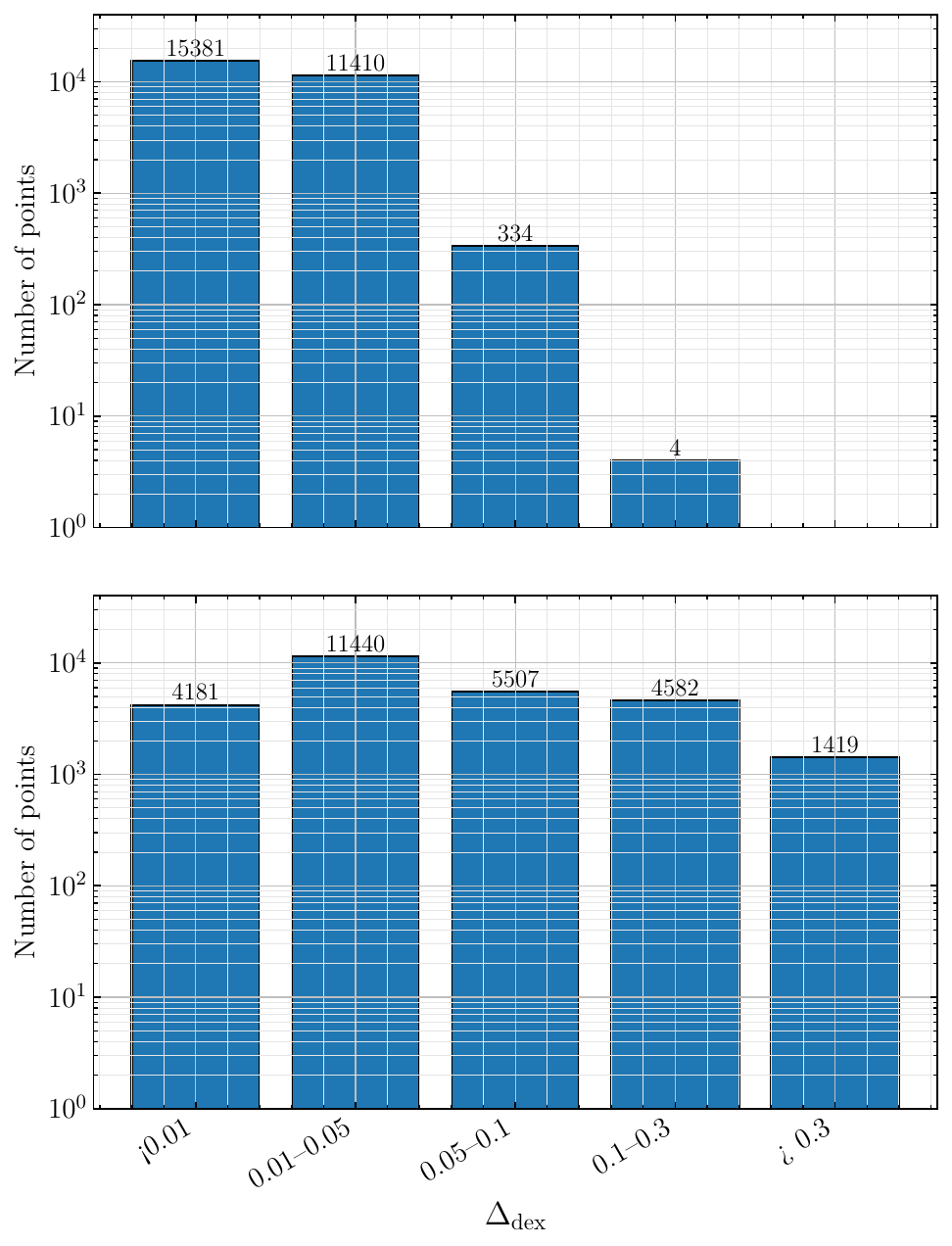}
    \caption{\textit{Top --} Error distribution histogram for the fiducial ML model. \textit{Bottom --} Error distribution histogram for trilinear interpolation over the force multiplier grid.}
    \label{fig:Trilinear}
\end{figure}

To benchmark the accuracy of the fiducial ML model, we compare its performance against trilinear interpolation over the original force multiplier grid. For each point in the testing set, we compare the corresponding \textsc{XSTAR} value to two approximate values: one obtained from the fiducial ML model trained on $N=80k$ points, and one obtained by trilinear interpolation over the 8 nearest neighbors of the test point. This comparison is designed to assess the accuracy of two approaches that can be used to estimate force multiplier values at arbitrary points in the hydrodynamic simulation.

As before, we omit points with $\log_{10} M_{\rm total} < -2$, since these make negligible contributions to the radiation force in the applications considered here. We also omit points along the boundary of the parameter space, where trilinear interpolation is poorly constrained. Figure~\ref{fig:Trilinear} shows that trilinear interpolation performs substantially worse than the fiducial ML model. Whereas the ML model gives $\Delta_{\rm dex} \gtrsim 0.05$ for only 1.2\% of test points, the trilinear interpolation scheme gives errors above this threshold for $\gtrsim 40\%$ of test points.

The poorer performance of trilinear interpolation is expected because it cannot resolve curvature in the force multiplier function between grid points. These results suggest that the accuracy of the photoionization treatment used in hydrodynamic simulations can be improved by using ML-trained models to generate higher-resolution lookup tables offline. The hydrodynamic calculation can then continue to use standard table interpolation during runtime, but on a denser grid sampled from the ML surrogate. This approach avoids the need to call the ML model directly inside the hydrodynamics code and should add minimal runtime cost, although at the expense of increased table size. In this test, the ML model was trained using $\lesssim 50\%$ of the available photoionization data, suggesting that the existing grid contains sufficient information to construct more accurate interpolated tables.

\section{Hydrodynamics}
\label{sec:hydro}

As an application of the ML force multiplier model, we study spherically symmetric line-driven winds launched from a star. This setup is intended as a controlled test problem for coupling the ML-based force multiplier prescription to hydrodynamic simulations, rather than as a detailed model of a particular observed system. The wind is irradiated by a stellar SED, which we initially take to be a blackbody. To explore the effect of spectral evolution through the flow, we allow the gas to have band-dependent absorption opacities in the range
$
0 \leq \kappa_a/\kappa_{\rm es} \leq 3.
$

Because the absorption opacity can differ between frequency bands, the shape of the radiation field changes as photons propagate through the wind. In particular, when the UV opacity is larger than the X-ray opacity, UV photons are preferentially absorbed and the radiation field hardens with radius. This provides a simple test of whether using local spectral information in the force multiplier calculation can alter the resulting wind dynamics.

\subsection{Photoionization via Machine Learning}

Previous models have used pre-tabulated photoionization data to include heating and cooling rates \citep[e.g.,][]{Dyda17} and line-force multipliers \citep[e.g.,][]{Dannen20}. The force multiplier prescription in \cite{2025arXiv250400117D} depended on the optical depth parameter and the temperature characterizing the irradiating blackbody spectrum. In this work, we extend this approach in two significant ways. First, we allow the force multiplier to depend simultaneously on three paramerters: the ionization parameter, the optical depth parameter and the temperature of the incident blackbody SED. Second, rather than interpolating over precomputed lookup tables, we directly query a random forest model to compute the force multiplier.

We train the ML model on blackbody spectra, whose shape is specified by a single temperature, \(T_{\rm bb}\). During the hydrodynamic calculation, we assign the local radiation field an equivalent blackbody temperature by matching its UV flux fraction, \(f_{\rm UV}\), using a precomputed lookup table. Frequency-dependent absorption generally produces a non-blackbody spectrum, so this procedure approximates the local spectral shape by a member of the blackbody training family. We focus on blackbody spectra, which provide a controlled framework for studying how changes in spectral shape affect the line force because the SED can be parameterized by a single quantity, the blackbody temperature $T_{\rm bb}$.  The resulting force multiplier therefore responds to changes in the local UV fraction, but does not capture spectral distortions that cannot be represented by a change in blackbody temperature.

To incorporate the photoionization calculations into the hydrodynamic simulations, we export the 
%Python-trained 
trained model to a format that can be read by our modified \textsc{Athena++} problem generator.  \textsc{Athena++} calls the exported ML model during runtime by passing the local gas parameters $(\xi,t,T_{\rm bb})$ to the model, evaluating the ensemble of trees, and returning the predicted force multipliers used to compute the line acceleration.

\subsection{Initial Conditions}

We choose parameters relevant to O stars. We take the mass to be $M = 50 M_{\odot}$ and radius $r_* = 10 r_{\odot}$. We set the Eddington fraction $\Gamma = 0.1$. We consider gas temperatures $10^{4} K \leq T \leq 10^{6} K$. For these temperatures, thermal driving is negligible \citep{Stone2009}.  

The simulation region extends radially from $r_* < r < 10 r_*$ This dynamical range allows the wind to fully accelerate and reach its terminal velocity. We use a logarithmically spaced grid of $N_r$ = 128 points and a scale factor $a_r = 1.008$ that defines the grid spacing recursively via $dr_{n+1} = a_r dr_{n}$. This resolution ensures that we resolve the acceleration of the wind through the critical point. 

At the inner boundary, we impose outflow boundary conditions on $v$ and $E$ while keeping the density fixed at $\rho_* = 10^{-10}$ g cm$^{-3}$ in the first active zone. This density ensures that we are sufficiently resolving the atmosphere at the base of the wind, but also providing sufficient mass to launch a wind given our choice of Eddington parameter. The continuum intensity of the radially outgoing rays is set by the Eddington parameter while the non-radial rays are set to zero. The input radiation spectra is set by choosing a blackbody temperature $T_{bb}$, which we use to determine the radiation fraction $f_i$ in each band.

At the outer boundary, we impose outflow boundary conditions on $\rho$, $v$ and $E$ and vacuum conditions on the continuum.

\subsection{Radiation Driven Wind}

\begin{figure}
    \centering
        \includegraphics[width=\linewidth]{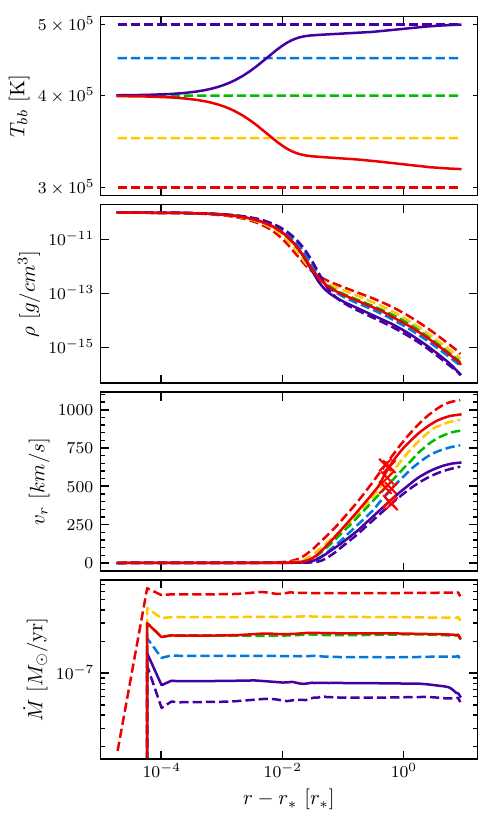}
    \caption{Radial profiles for irradiating SED tamperautre $T_{\rm{bb}}$, density $\rho$, velocity $v_r$ and mass flux $\dot{M}$. Models with dashed lines have constant $T_{\rm{bb}}$ whereas solid lines indicate models with radially dependent SED. The color of these latter models is chosen to match model with approximately the same temperature at large radii.}
    \label{fig:ModelComparison}
\end{figure}

To test the codes capability to handle changing SEDs, we consider a set of wind solutions where 1) the irradiating SEDs blackbody temperature is constant throughout the flow 2) the local $T_{\rm{bb}}$ is changing. Practically, we achieve these position dependent SED models by making one band optically thin, and one band optically thick with an absorption opacity $\kappa_{a,\nu} = 3 \kappa_{es}$. For the model with $\kappa_{UV} > 0$, the UV fraction \emph{decreases} with radius, hence $T_{\rm{bb}}$ increases with radius. Likewise, $\kappa_{X} > 0$ causes $T_{\rm{bb}}$ to decrease with radius.

We show the steady-state solutions for constant (dashed lines) and positionally varying (solid lines) $T_{\rm{bb}}$ in Fig \ref{fig:ModelComparison}. The top panel shows the temperature of the irradiating SED. For models with constant $T_{\rm{bb}}$, the colors indicate the irradiating SED's temperture throughout the flow - $T_{\rm{bb}} =$ $3 \times 10^5$ (red), $3.5 \times 10^5$ (orange), $4 \times 10^5$ (green), $4.5 \times 10^5$ (cyan) and $5 \times 10^5$ (blue). For models with position varying spectra, $T_{\rm{bb}} =$ $4 \times 10^5$ at the base of the wind and the color is chosen to approximately match the SED temperature at large radii, namely $3 \times 10^5$ for the X-ray absorption (red) and $5 \times 10^5$ UV absorption model (blue) respectively. 

The bottom panel shows the mass flux, which is approximately independent of radius, consistent with a steady state. We see $\dot{M}$ increases as a function of decreasing $T_{\rm{bb}}$, as expected for a force multiplier decreasing as the spectrum hardens. Likewise, the velocity of the outflows decreases with the hardening of the SED, consistent with the decreasing force multiplier. 

We can use the constant SED solutions to get an intuition for what happens when the SED is position dependent. We compare to both the constant $T_{\rm{bb}} = 4 \times 10^{5}$ K model to the UV (blue) and X-ray (red) absorption models. 

The UV absorption model is well described by the $T_{\rm{bb}} = 5 \times 10^{5}$ K (blue model), with both the mass flux and velocity profile matching the constant SED model. This is consistent with the mass flux being set by the driving force at the base of the flow (at the critical point, indicated by the red crosses near $r - r_* \sim 0.5 r_*$) and the velocity being increased with additional forces further out in the winds \cite{1999isw..book.....L}.Conversely, in the X-ray absorption case (red line), the mass flux matches the $T_{\rm{bb}} = 4 \times 10^{5}$ K model, but the velocity increases by $\sim 20\%$. If the mass flux were being set by the SED at the critical point, we would expect a value intermediate to the constant red and orange models. For such non-trivial models, where SED is varying with position one cannot rely exclusively on intuition from ideal analytic models. These results suggest that changes to the force multiplier, as a result of a varying SED, can have significant effects on the resulting wind. As the physics of these flows is entirely determined by the radiation force, accurately computing the SED and its resultant force multiplier is crucial for consistent, line driven wind solutions. 

\section{Discussion}
\label{sec:discussion}

We have developed and tested a method of computing force multipliers in radiation hydrodynamic simulations using a ML model trained on photoionization data. We have applied this method to the problem of a spherically symmetric line driven wind from a star. We find that the key hydrodynamic properties of the wind - mass flux and outflow velocity, are altered when the irradiating SED varies with position. These results highlight the importance of using self-consistent radiation fields in both the photoionization and rad-hydro calculations.

We adopted Random Forest regression to predict force multipliers after comparing its performance with Gradient Boosting and MLP models. Random Forest achieved the lowest errors among the configurations tested for blackbody SEDs. However, this performance does not establish a universal preference for Random Forest. Other methods may perform better for different physical problems or microphysics. The exact model selection should therefore be thoroughly tested for the desired application. A related question is what variables to use to train the model. In this work we use the force multiplier but other, equally physically motivated parameters such as the Gayley $\bar{Q}$ parametrization of the line force \citep{Gayley1995} may be used.

We have applied this method to a very specific set of blackbody SEDs, which can be parameterized with a single parameter. Firstly, we used a two band approximation and the fit can be improved my using multiple bands for the spectral fit (see for example \cite{Stanley2026}). Further, one would like to relax the assumption that the irradiating SED is apriori blackbody and extend this same ML technique to more general spectra. The key problem that must be solved is how to best parametrize the SED and what part of this parameter space should be used to train the model. Consider the problem of an AGN disc wind. Assuming a simple model for the radiation field such as a Shakura-Sunyaev thin disc and a power-law corona, we can parametrize the SED with black hole mass, mass accretion rate, corona power-law index and position (5 parameters \cite{Smith24}). One can uniformly sample this parameter space to generate the training/test set. An important question is how large the training set needs to be to achieve satisfactory accuracy. A separate, but related question is whether the hydro simulation can be run with the number of frequency bands required by the ML model. These may very well be problem dependent and may ultimately determine if this method is suitable. 

Once trained, a ML force multiplier model can be used with any hydrodynamics code and radiation transfer method. We demonstrate the method with \textsc{Athena++} which solves the radiation field along rays by explicitly solving the time dependent radiation transfer equation. However, the ML model only requires as input the local SED, which can be computed from any number of methods such as ML, Monte Carlo, etc... For example, our method is complementary to the method described in the AGN disc wind study \cite{Scepi2026}. They use \textsc{PLUTO} \citep{PlutoMethod} to solve the hydrodynamics equation, \textsc{SIROCCO} \citep{Matthews2025} to evolve the radiation field and 
% finally an extensive ($\sim$ 450 000 line) list \cite{Parkin2013}
finally used an external line list with $\approx$ 450,000 lines to f
to compute the force multiplier from the local SED and ionization state. This final step could be replaced with a ML model trained on SEDs relevant to the problem at hand such as CVs \cite{Higginbottom2024,Mosallanezhad2025} or AGN \cite{Scepi2026}.

We show that ML models perform significantly better than trilinear interpolation for interpolating force multiplier values. Notably, force multiplier values have $\Delta_{\rm{dex}}\leq 0.05$ errors for $\lesssim 1\%$ of points in the phase space. This was achieved with $N = 80k$ training points, less than half as many points in the full data set available. This suggests that one can improve accuracy of interpolated data in the hydrodynamics model by training a ML model and using that model to generate much finer resolution grids which can then be queried in the hydrodynmics code. Thus, a ML approach may be useful, even if it is not incorporated in a hydrodynamics model directly and one still relies on precomputed tables.   

The approach presented here can be extended to other microphysical calculations, such as opacities heating/cooling rates, chemical or nuclear networks, whose outputs can be determined from a specified local gas state and radiation field (see references in Section \ref{sec:introduction}). These calculations can be performed independently to train a surrogate, while their feedback on the dynamics is retained through runtime coupling. This separation does not directly apply to unresolved turbulence for example, where interactions across physical scales require a dynamical closure rather than an independently tabulated microphysical response. The broader applicability of our approach therefore depends on identifying a sufficiently complete set of local inputs that captures the relevant microphysics. ML may thus provide a new tool to couple a system's large scale dynamics and the applicable microphysics.

\appendix

\section{Force Multipliers}
\label{sec:appendix}

\begin{figure*}
    \centering       
        \includegraphics[scale=0.48]{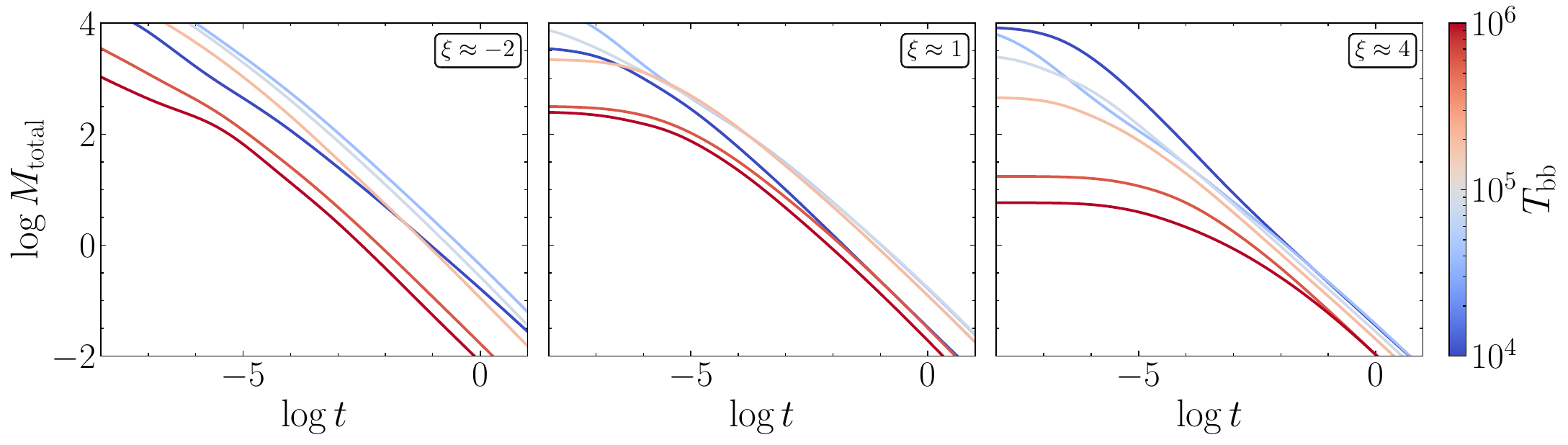} 
    \caption{Force multiplier as a function of optical depth parameter for a representative sample of $\log \xi = -2, 1, 4$ and $T_{bb} = 10^4, 4 \times 10^4, 8 \times 10^4, 2 \times 10^5, 6 \times 10^5, 10^6$ K}
    \label{fig:Mtotal}
\end{figure*}

\begin{figure*}
    \centering       
        \includegraphics[scale=0.36]{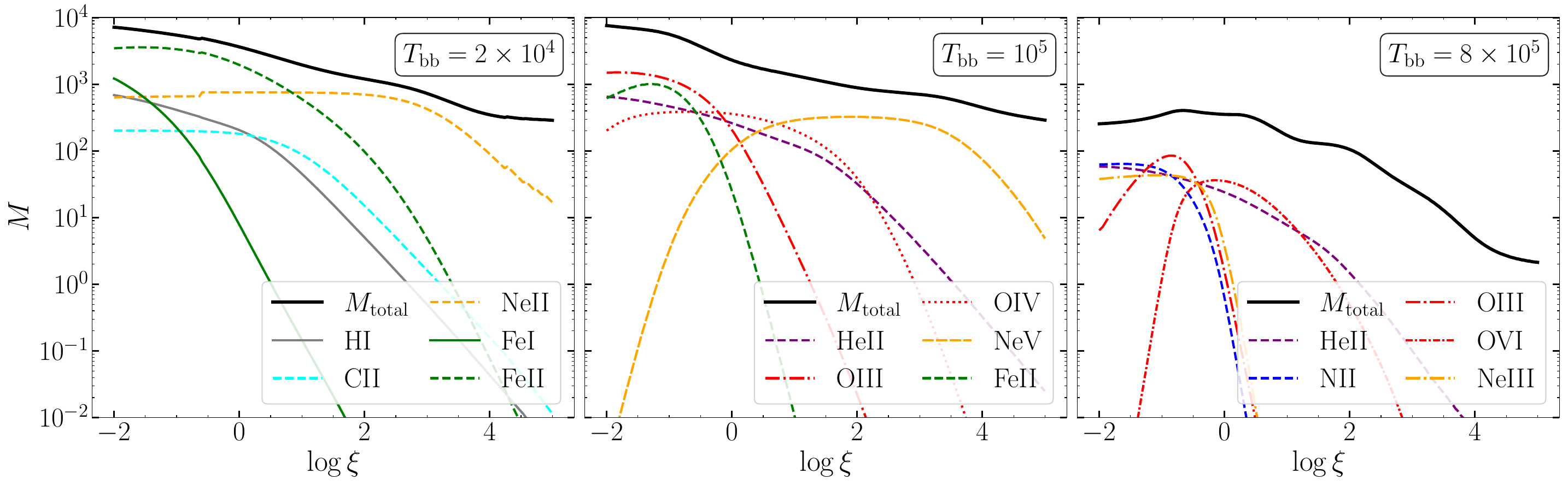} 
    \caption{Force multiplier as a function of ionization parameter and the respective contributions of the most dominant ions.}
    \label{fig:IonContributions}
\end{figure*}

\begin{figure*}
    \centering       
        \includegraphics[scale=0.36]{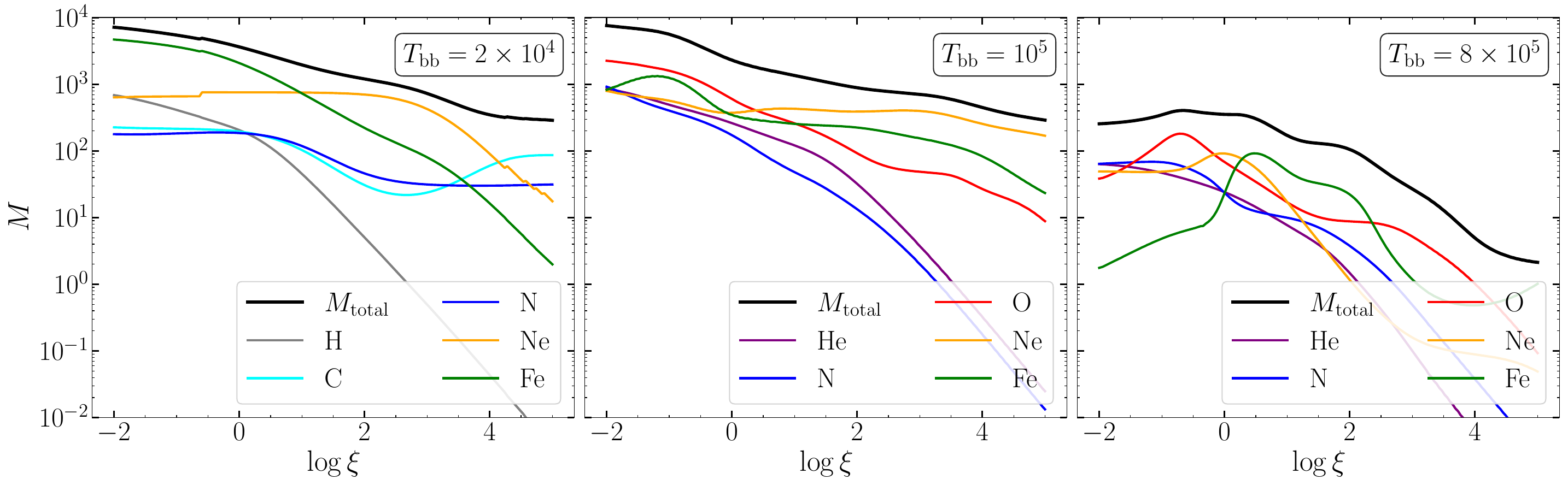} 
    \caption{Force multiplier as a function of ionization parameter and the respective contributions of the most dominant elements.}
    \label{fig:ElementContributions}
\end{figure*}

\begin{table*}
\centering
\small
\setlength{\tabcolsep}{4pt}
\begin{tabular}{c|cc||cc||cc||cc||cc}
\hline\hline
& \multicolumn{10}{c}{$T_{\rm{bb}}$} \\

 & \multicolumn{2}{c||}{$2\times10^4\,\mathrm{K}$} & 
   \multicolumn{2}{c||}{$6\times10^4\,\mathrm{K}$} & 
   \multicolumn{2}{c||}{$1\times10^5\,\mathrm{K}$} & 
   \multicolumn{2}{c||}{$4\times10^5\,\mathrm{K}$} &
   \multicolumn{2}{c}{$8\times10^5\,\mathrm{K}$} \\
\hline\hline

\multirow{7}{*}{\rotatebox{90}{$\log \xi \simeq -2$}}
& \multicolumn{2}{c||}{$M_{\rm total}=7178$}
& \multicolumn{2}{c||}{$M_{\rm total}=9136$}
& \multicolumn{2}{c||}{$M_{\rm total}=7601$}
& \multicolumn{2}{c||}{$M_{\rm total}=1118$}
& \multicolumn{2}{c}{$M_{\rm total}=256$} \\

& Ion & $M_L$ & Ion & $M_L$ & Ion & $M_L$ & Ion & $M_L$ & Ion & $M_L$ \\
\cline{2-11}
& \ion{Fe}{2} & 3464 & \ion{Fe}{2} & 1635 & \ion{O}{3}  & 1496 & \ion{He}{2} & 276 & \ion{N}{2}  & 63 \\
& \ion{Fe}{1} & 1219 & \ion{H}{1}  & 1386 & \ion{H}{1}  & 660  & \ion{Ne}{3} & 208 & \ion{He}{2} & 58 \\
& \ion{H}{1}  & 688  & \ion{N}{2}  & 854  & \ion{He}{2} & 651  & \ion{O}{3}  & 187 & \ion{Ne}{3} & 38 \\
& \ion{Ne}{2} & 634  & \ion{O}{3}  & 833  & \ion{Fe}{2} & 618  & \ion{O}{2}  & 63  & \ion{O}{1}  & 20 \\
& \ion{Si}{2} & 211  & \ion{C}{2}  & 636  & \ion{N}{2}  & 551  & \ion{N}{2}  & 59  & \ion{O}{2}  & 12 \\

\hline \hline

\multirow{7}{*}{\rotatebox{90}{$\log \xi \simeq 0$}}
& \multicolumn{2}{c||}{$M_{\rm total}=3644$}
& \multicolumn{2}{c||}{$M_{\rm total}=3338$}
& \multicolumn{2}{c||}{$M_{\rm total}=2312$}
& \multicolumn{2}{c||}{$M_{\rm total}=1085$}
& \multicolumn{2}{c}{$M_{\rm total}=353$} \\

& Ion & $M_L$ & Ion & $M_L$ & Ion & $M_L$ & Ion & $M_L$ & Ion & $M_L$ \\
\cline{2-11}
& \ion{Fe}{2} & 1956 & \ion{O}{3}  & 778 & \ion{O}{4}  & 360 & \ion{Fe}{7} & 174 & \ion{Ne}{5} & 40 \\
& \ion{Ne}{2} & 758  & \ion{Fe}{3} & 402 & \ion{He}{2} & 262 & \ion{Fe}{6} & 167 & \ion{O}{6}  & 35 \\
& \ion{H}{1}  & 206  & \ion{Fe}{2} & 280 & \ion{O}{3}  & 208 & \ion{Ne}{5} & 162 & \ion{Ne}{6} & 33 \\
& \ion{C}{2}  & 181  & \ion{H}{1}  & 252 & \ion{Ne}{4} & 179 & \ion{Ne}{6} & 122 & \ion{C}{4}  & 31 \\
& \ion{N}{2}  & 158  & \ion{N}{3}  & 233 & \ion{N}{3}  & 121 & \ion{He}{2} & 73  & \ion{He}{2} & 24 \\

\hline \hline

\multirow{7}{*}{\rotatebox{90}{$\log \xi \simeq 2$}}
& \multicolumn{2}{c||}{$M_{\rm total}=1207$}
& \multicolumn{2}{c||}{$M_{\rm total}=889$}
& \multicolumn{2}{c||}{$M_{\rm total}=890$}
& \multicolumn{2}{c||}{$M_{\rm total}=200$}
& \multicolumn{2}{c}{$M_{\rm total}=106$} \\

& Ion & $M_L$ & Ion & $M_L$ & Ion & $M_L$ & Ion & $M_L$ & Ion & $M_L$ \\
\cline{2-11}
& \ion{Ne}{2} & 703 & \ion{O}{4}  & 175 & \ion{Ne}{5} & 325 & \ion{Fe}{10} & 48 & \ion{Si}{10} & 27 \\
& \ion{Fe}{3} & 124 & \ion{Ne}{4} & 129 & \ion{Fe}{6} & 139 & \ion{Fe}{11} & 43 & \ion{S}{11}  & 15 \\
& \ion{Fe}{2} & 99  & \ion{Fe}{5} & 86  & \ion{Fe}{7} & 80  & \ion{Si}{9}  & 15 & \ion{Fe}{12} & 9 \\
& \ion{O}{3}  & 55  & \ion{Ar}{6} & 76  & \ion{O}{4}  & 40  & \ion{Fe}{12} & 14 & \ion{Fe}{11} & 6 \\
& \ion{S}{4}  & 36  & \ion{C}{4}  & 38  & \ion{Mg}{5} & 37  & \ion{Fe}{9}  & 10 & \ion{Si}{9}  & 6 \\

\hline \hline

\multirow{7}{*}{\rotatebox{90}{$\log \xi \simeq 4$}}
& \multicolumn{2}{c||}{$M_{\rm total}=351$}
& \multicolumn{2}{c||}{$M_{\rm total}=627$}
& \multicolumn{2}{c||}{$M_{\rm total}=453$}
& \multicolumn{2}{c||}{$M_{\rm total}=76$}
& \multicolumn{2}{c}{$M_{\rm total}=5$} \\

& Ion & $M_L$ & Ion & $M_L$ & Ion & $M_L$ & Ion & $M_L$ & Ion & $M_L$ \\
\cline{2-11}
& \ion{Ne}{2} & 93 & \ion{Ne}{5}  & 256 & \ion{Ne}{6}  & 172 & \ion{Si}{10} & 18 & \ion{O}{8}  & 1 \\
& \ion{O}{3}  & 55 & \ion{Ne}{4}  & 66  & \ion{Ne}{5}  & 77  & \ion{S}{11}  & 12 & \ion{S}{13}  & 1 \\
& \ion{C}{4}  & 44 & \ion{O}{4}   & 66  & \ion{Fe}{7}  & 58  & \ion{Fe}{13} & 10 & \ion{Ar}{14} & 1 \\
& \ion{S}{4}  & 44 & \ion{O}{6}   & 41  & \ion{Mg}{6}  & 27  & \ion{Fe}{14} & 8  & \ion{S}{12}  & 1 \\
& \ion{Si}{4} & 32 & \ion{O}{5}   & 36  & \ion{Fe}{8}  & 24  & \ion{Fe}{12} & 5  &              &   \\

\hline\hline
\end{tabular}
\label{tab:ion}
\caption{Dominant ionic contributions to the force multiplier $M_{\rm total}$ across a range of ionization parameters and blackbody temperatures for optically thin, $\log t = -6$, gas. The table highlights the shift from low-ionization species at small $\xi$ to highly ionized metals at large $\xi$, accompanied by a decline in $M_{\rm total}$.}
\end{table*}

Fig \ref{fig:Mtotal} shows how the total force multiplier varies as a function of optical depth parameter for different irradiating blackbody SEDs and some representative ionization states $\log \xi$ = -2, 1 and 4. For all ionization parameters and blackbody SEDs $M_{\rm{total}}$ declines smoothly with increasing $t$, reflecting the reduced effectiveness of line driving as lines become optically thick. For small optical depth parameter and high ionization, the strength of the line driving remains relatively constant for the harder SEDs. At large $t$, the force multiplier approaches an approximately power-law form, $M(t) \propto t^{-\alpha}$, analogous to the CAK force multiplier prescription. The slopes in Fig \ref{fig:Mtotal} are typically $0.6 \leq \alpha \leq 0.8$, consistent with a CAK type prescription, though the detailed curvature and normalization depends on $T_{\rm{bb}}$. At low ionization $\xi = -2$, softer spectra (blue curves) produce larger force multipliers, while harder spectra (red curves) are less efficient. As ionization increases, the overall magnitude of $M_{\rm{total}}$ decreases and the spread between different blackbody SEDs increases, indicating that as the gas becomes more highly ionized it is more sensitive to the detailed spectral shape.

Fig \ref{fig:IonContributions} shows the total force multiplier $M_{\rm{total}}$ as a function of ionization parameter (thick black line), together with the five most dominant ionic contributors at fixed $\log t = -6$ for several representative $T_{bb} = 2 \times 10^4 \ \rm{K}, \ 1 \times 10^5 \ \rm{K} \ \rm{and} \ 8 \times 10^5 \ \rm{K}$. Each element is plotted with a separate color and ionizations states are indicated via the line style. At low $\xi$, the force is dominated by low-ionization species, which provide large line opacities and therefore strong momentum coupling. As $\xi$ increases, these species are progressively ionized away and the contribution shifts to higher ionization states, typically from lighter elements first and then heavier species such as Fe-group ions at the highest $\xi$ and hottest blackbody temperatures. This produces a characteristic decline in $M_{\rm{total}}$ with increasing ionization, along with a changing hierarchy of contributing ions.

To further clarify these trends, Fig.~\ref{fig:ElementContributions} shows the same calculation with contributions summed over all ionization states of each element. This representation isolates how the $M_{\rm{total}}$ opacity budget is partitioned across elements, independent of detailed ionization structure. At low $\xi$, the force is shared among several abundant light elements, with C, N, and Ne contributing at comparable levels alongside Fe. As $\xi$ increases, the contribution from hydrogen drops rapidly, reflecting its limited line opacity, while metals dominate the force multiplier. The relative importance of individual elements evolves with both $\xi$ and $T_{\rm bb}$: for cooler spectra, Fe provides a substantial fraction of the opacity over a wide range of $\xi$, whereas at higher temperatures and ionization parameters, intermediate-mass elements such as O and Ne become more prominent before eventually declining as they are fully stripped. At the highest $\xi$, all elemental contributions decrease sharply, leading to the overall decline in $M_{\rm total}$. This element-summed view reinforces the picture that the reduction in line driving efficiency arises from both the progressive ionization of individual species and a redistribution of opacity across elements with differing line densities and ionization thresholds.

Table \ref{tab:ion} summarizes the dominant ionic contributors to the force multiplier across a range of ionization parameters and blackbody temperatures. A clear and systematic trend is observed with increasing ionization parameter. At low ionization ($\log \xi \simeq -2$), the force multiplier is dominated by low-ionization species, particularly Fe~II, Fe~I, and H~I. For example, at $T_{\rm bb} = 2\times10^4\,\mathrm{K}$, Fe~II alone contributes $M_L \simeq 3464$, followed by Fe~I (1219) and H~I (688), indicating that line driving is primarily mediated by abundant, weakly ionized metals and neutral hydrogen in this regime. As the ionization parameter increases to $\log \xi \simeq 0$, these low-ionization species remain important but their relative contributions decline, and moderately ionized species such as O~III, Ne~II, and Fe~III begin to play a more significant role.

At higher ionization ($\log \xi \simeq 2$), the dominant contributors shift to intermediate and high-ionization states, with species such as Ne~V, Fe~VI, and O~IV becoming prominent. This transition is accompanied by a substantial reduction in the total force multiplier. For instance, at $T_{\rm bb} = 2\times10^4\,\mathrm{K}$, $M_{\rm total}$ decreases from 7178 at $\log \xi \simeq -2$ to 1207 at $\log \xi \simeq 2$. By $\log \xi \simeq 4$, the gas is highly ionized and only a small number of highly ionized species contribute, such as O~VIII, S~XIII, and Ar~XIV, with individual contributions of order unity. Correspondingly, the total force multiplier drops to very low values (e.g., $M_{\rm total} \simeq 5$ at $T_{\rm bb} = 8\times10^5\,\mathrm{K}$), reflecting the depletion of bound--bound opacity as atoms become fully stripped.

The dependence on spectral shape is also evident. At fixed ionization parameter, increasing $T_{\rm bb}$ shifts the dominant ions toward higher ionization states. For example, at $\log \xi \simeq 2$, the dominant contributors evolve from Ne~II and Fe~III at $T_{\rm bb} = 2\times10^4\,\mathrm{K}$ to Ne~V and Fe~VI at $T_{\rm bb} = 10^5\,\mathrm{K}$, and further to Fe~X and Fe~XI at $T_{\rm bb} = 4\times10^5\,\mathrm{K}$. This reflects the harder radiation field associated with higher blackbody temperatures, which increases the ionization state of the gas. Additionally, hydrogen is only a significant contributor at low ionization and becomes negligible at higher $\xi$, consistent with its limited line opacity and rapid ionization.

Overall, these trends demonstrate that the efficiency of line driving is governed by a combination of ionization balance and spectral hardness. Low-ionization gas with a rich population of partially ionized metals produces large force multipliers, while highly ionized gas exhibits a marked decline in line driving efficiency due to the loss of available line transitions.

\section*{Acknowledgments} 
Support for this work was provided by the National Aeronautics and Space Administration under TCAN grant 80NSSC21K0496. We thank the entire DAWN TCAN collaboration for fruitful discussions.  The authors acknowledge Research Computing at The University of Virginia for providing computational resources and technical support that have contributed to the results reported within this publication.

\section*{Data Availability Statement}
The simulations were performed with the publicly available code \textsc{Athena++} available at https://github.com/PrincetonUniversity/athena and \textsc{XSTAR} available at https://heasarc.gsfc.nasa.gov/xstar/xstar.html The authors will provide the force multiplier tables and machine learning models to interested parties. The authors will also provide any additional problem generators and input files upon request.

%%%%%%%%%%%%%%%%%%%%%%%%%%%%%%%%%%%%%%%%%%%

\bibliographystyle{mnras}
\bibliography{progalab-shared}

%\bsp	% typesetting comment
\label{lastpage}

\end{document}